\documentclass[sigconf]{acmart}

\usepackage{graphics} 
\usepackage{multirow}
\usepackage{dirtytalk}
\usepackage{booktabs}
\usepackage[section]{placeins}

\long\def\omitit#1{}
\DeclareGraphicsExtensions{.eps,.ps,.png,.pdf}

\begin{document}

\title{Generative AI in the Software Engineering Domain: Tensions of Occupational Identity and Patterns of Identity Protection}



\author{Anuschka Schmitt }
\email{a.schmitt2@lse.ac.uk}
\affiliation{%
 \institution{The London School of Economics and Political Science}
 \city{London}
 \country{United Kingdom}
}

\author{Krzysztof Z. Gajos}
\orcid{0000-0002-1897-9048}
\email{kgajos@seas.harvard.edu}
\affiliation{%
 \institution{Harvard University}
 \city{Cambridge}
 \country{USA}
}

\author{Osnat Mokryn}
\email{omokryn@is.haifa.ac.il}
\affiliation{%
 \institution{University of Haifa}
 \country{Israel}
}



\begin{abstract}
The adoption of generative Artificial Intelligence (GAI) in organizational settings calls into question workers' roles, and relatedly, the implications for their long-term skill development and domain expertise. In our qualitative study in the software engineering domain, we build on the theoretical lenses of occupational identity and self-determination theory to understand how and why software engineers make sense of GAI for their work. We find that engineers' sense-making is contingent on domain expertise, as juniors and seniors felt their needs for competence, autonomy, and relatedness to be differently impacted by GAI. We shed light on the importance of the individual's role in preserving tacit domain knowledge as engineers engaged in sense-making that protected their occupational identity. We illustrate how organizations play an active role in shaping workers' sense-making process and propose design guidelines on how organizations and system designers can facilitate the impact of technological change on workers' occupational identity.
\end{abstract}



\maketitle

\section{INTRODUCTION}
The widespread availability of Generative Artificial Intelligence (GAI)-based models capable of performing highly complex tasks is expected to elicit a paradigm shift in the workforce~\cite{frank2019toward, agrawal2023we, eloundou2023gpts}. 
However, the exact nature of GAI's impact on the workforce is under debate. While some predictions go as far as suggesting the replacement of knowledge-intensive jobs in the near future ~\cite{kanbach2024genai,gs2024labormarket}, GAI also offers the potential to promote equality and enable unparalleled productivity gains by complementing workers' skills and expertise~\cite{acemoglu2023can,agrawal2023we,girotra2023ideas}. 
Empirical research found the augmentation of human work through technology to increase efficiency and productivity, e.g., for taxi drivers optimizing their route planning with AI~\cite{kanazawa2022ai}, or novice customer support agents handling customer queries~\cite{brynjolfsson2023generative}. Especially for knowledge-intensive work, GAI has become relevant, e.g., in terms of saving time for creative~\cite{girotra2023ideas}, legal~\cite{katz2023gpt}, and consulting tasks~\cite{DellAcqua2023-ia}. However, predicted productivity gains and workforce replacements introduce novel challenges around learning and the preservation of tacit domain knowledge in the workplace.

Scholars in Human-Computer Interaction (HCI) have emphasized the importance of considering worker perspectives in contexts prone to technological changes~\cite{karusala2023speculating,sergeeva2023postphenomenological}. As GAI increasingly defines, manages, or even replaces what workers do, it changes not only the nature of work processes and individual tasks. More so, it disrupts how workers make sense of their work and how they view themselves. Taking into perspective the individual perceptions of workers becomes particularly important in consideration of `the last mile problem': while performance gains through novel technology are in theory possible, the adoption and appropriate use of such technology hinge on the individual~\cite{Anthony2021-gr, Beane2019-wi}.

However, how workers perceive GAI to impact their role and work is not fully understood. A recent study with young professionals showed that some felt that working with GAI enhanced their sense of competence and autonomy, while others experienced a diminished sense of ownership and perceived a lack of challenge~\cite{kobiella2024if}. Leveraging the theoretical lenses of occupational identity and self-determination theory (SDT) enables us to explore the impact of GAI in the workplace from a fresh yet crucial perspective. Through their occupational identity, workers make sense of their occupation and their own role, providing workers with meaning and a sense of distinctiveness~\cite{Alvesson2002-zr, Vaast_undated-ka}. According to SDT, satisfying the three psychological needs for competence, autonomy, and relatedness, is key to workers' sense-making process and intrinsic motivation~\cite{deci2012self, deci2017self, Ryan2000-vv, Van_den_Broeck2021-hq}. While extant findings shed light on the ontological feasibility of augmenting human work through GAI, it is unclear how this paradigm shift is experienced and made sense of by affected workers. In this study, we explore the effect of GAI on software engineers' sense-making of their occupational identity by addressing the following two research questions (RQ): 

\textit{RQ1: How does GAI affect software engineers' need for competence, autonomy, and relatedness?} 

\textit{RQ2: Which underlying individual and organizational factors can help explain software engineers' sense-making of GAI?} 

To address our research questions, we conducted a qualitative study of engineers in a middle-sized software organization in the United States. The context of our study enables us to explore a work domain where the use of GAI, theoretically, can influence the nature and outcome of tasks and occupations, yet where the workers currently choose whether to use GAI or not. Leveraging survey data and semi-structured interviews, we study engineers' {\em sense-making of GAI for their occupational identity, and its impact on work-related psychological needs}. 

We show that software engineers varied in their reactions to GAI. 
We consider software engineers' varying levels of domain expertise as an important factor in shaping engineers' sense-making process of GAI. 
We find that (i) GAI impacts software engineers' need for competence, autonomy, and relatedness and that (ii) threats toward these three needs are experienced differently by junior and senior software engineers. In response to these threats, we unpack prevalent patterns of how software engineers protect their existing occupational identity, and thereby also preserve their skill development. Lastly, we identify how organizational measures and latent messages by management implicitly moderate workers' sense-making process of their occupational identity.

Addressing our research questions is relevant for three key reasons. First, we contribute to occupational identity literature~\cite{Bechky2011-dc, Sveningsson2003-vr, Vaast2021-ll} by shedding light on how GAI as a new technology impacts software engineers' occupational identity. If highly skilled knowledge work is prone to automation and augmentation, it is crucial to understand how knowledge workers themselves make sense of the advent of GAI for their occupation. Our findings challenge prevalent assumptions of how the introduction of GAI in the workplace will lead to a de-skilling of knowledge workers and a loss of tacit domain expertise. We shed light on the importance of the individual's role in preserving tacit domain knowledge as workers engage in sense-making that protects their occupational identity. Second, we show that workers' sense-making of their occupational identity is contingent on domain expertise. 
We find that juniors and seniors' psychological needs can help explain the underlying reasons for engineers' different reactions to GAI. Third, the introduction of GAI into work raises the question of how we should envision desirable and sustainable forms of technological change in the workplace, and AI augmentation, specifically~\cite{acemoglu2023can, agrawal2023we, Orlikowski1992-ga}. We contribute to this discussion by demonstrating the importance of the organization's role in sustaining knowledge workers' agency in the form of (i) enacting accountability and responsibility as a senior knowledge worker, and (ii) exercising ownership to ensure skill development as a junior knowledge worker.

\section{RELATED WORK}
\subsection{Knowledge Work, Occupational Identity and the Importance of Self-Determination}
A body of work in organizational and information systems studies, as well as in CSCW and HCI, has looked at the introduction of digital technology in the workplace and its effect on knowledge work, specifically~\cite{Anthony2021-gr, Gero2023-ro, Widder2021-kx, Karusala2021-pa, Sheehan2023-ba}. Knowledge work is distinguishable from other forms of work by requiring the application of theoretical, analytical, and tacit knowledge, and is thus highly contextualized and domain-specific~\cite{Allen2022-jn, autor_2015-ih, Dreyfus1984-rr, Polanyi2009-hh, Schultze2000-yc}. Programmers, analysts, and researchers are common examples of knowledge workers~\cite{Schultze2000-yc}.

\textit{Occupational identity} can be understood as a worker's identity and is ``the overlap of `who we are' and `what we do'~''~\cite{Nelson2014-kt}. People rely on their occupational identity to shape and communicate their understanding of what their occupation is and what its members do. Occupational identity delivers a sense of coherence and distinctiveness, often in relation to other occupations~\cite{Vaast2021-ll, Sveningsson2003-vr}. Several studies have noted the importance of knowledge work in shaping workers’ understanding of self, and in giving meaning to their work~\cite{Brown2015-tc, Ibarra2010-xe}. Next to positive implications for an individual's well-being, a strong occupational identity offers important organization-level implications. Beyond continuous learning and teaching on the job, productivity through knowledge work is contingent on a sense of feeling responsible for one’s contributions~\cite{Allen2022-jn}. This is where the importance of workers' self-determination comes into play.

According to \textit{self determination} theory (SDT), the satisfaction of the three psychological needs of competence, autonomy, and relatedness, is key to helping explain workers' intrinsic motivation and, relatedly, their well-being and further important work outcomes~\cite{deci2012self,deci2017self}. The \textit{need for competence} pertains to an individual's perceived efficacy of doing their work~\cite{Ryan2000-vv}. Workers' \textit{need for autonomy} is related to their desire to be ``agents of their own behavior'' rather than feeling externally steered~\cite{Orlikowski1992-ga}. 
Lastly, the \textit{need for relatedness} refers to workers' need to feel connected with colleagues, supervisors, and the organization as a whole~\cite{deci2017self}. SDT literature has explored how the fulfillment of these three needs has been associated with greater job satisfaction, organizational commitment, and more proactive efforts in crafting one's job~\cite{deci2012self,Ryan2000-vv, Ryan2007-rf, shen2024perceived, Van_den_Broeck2021-hq}. With the introduction of new digital technology in the workplace, however, workers' psychological needs cannot be satisfied consistently, yet may be challenged~\cite{Gagne2022-lg}. SDT has proven useful in understanding the \textit{why} of behavior related to identity work~\cite{Strachan2013SDTIdentity}. We therefore consider the needs for competence, autonomy, and relatedness in understanding and explaining software engineers' sense-making of their occupational identity in relation to GAI. We view our work as complementing and extending SDT: while the main model of this theory focuses on workers' motivation, we focus on its conceptualization and understanding of competence, autonomy, and relatedness as key to a worker's identity.

\subsection{Technology as a Trigger for Identity Work}
Shifting fulfillment of the key psychological needs of SDT and the resulting process of identity work can be conditioned by intra- and extra-organizational influences. As such, occupational identity is a dynamic and ever-evolving process~\cite{Bechky2011-dc, Sveningsson2003-vr, Vaast2021-ll}. Through \textit{identity work}, we change the meaning of who we are in the workplace. This can result in us maintaining or strengthening our existing occupational identity, but also forming and revising our occupational identity~\cite{Sveningsson2003-vr}.

Organizational influence can exacerbate or diminish how workers make sense of their occupational identity. An organization can unconsciously control how workers' identity work is ``triggered'' or even exercise control by explicitly triggering workers' identity work. This enactment of identity work can be referred to as ``modes of identity work regulation''~\cite{Alvesson2002-zr}. As such, organizations may purposefully induce identity work, e.g., through events or job promotions. But identity work can also occur unexpectedly, such as through the informal adoption of a new digital technology~\cite{Alvesson2002-zr}.

Literature has illustrated how technology can play a determining role in workers' occupational identity~\cite{Barrett1999-bv,Carter2015-ph, Lamb2005-dh, Stein2013-dt,Nyberg2009-ro, Orlikowski2008-wl} by serving as an identity referent~\cite{Ravasi2010-oy, Vaast2021-ll} or by being viewed as a form of an extended self~\cite{Nyberg2009-ro, Orlikowski2008-wl}. 
Technology might also threaten one’s occupational identity as it might render an occupation obsolete~\cite{autor2003outsourcing}, e.g., if a programming language a software engineer gained expertise in is replaced ~\cite{Vaast2021-ll}. Some studies also illustrate how digital technologies can lead to novel occupational identities as workers frame their sense of self in relation to the IT they are using or explicitly refraining from using~\cite{Stein2013-dt}. In a study exploring the impact of the internet~\cite{Nelson2014-kt}, workers used internet search for previous search practices and extended aspects of internet search to other practices, ultimately redefining their occupational identity by leveraging a technology that was claimed to replace them. However, some workers missed innovation opportunities because of their deep knowledge of non-internet research. They coined this phenomenon as a ``paradox of expertise'', positing that workers with rich domain expertise may not necessarily be best positioned to leverage new technologies. This raises important questions, such as whether identity work is contingent on domain experience.

Aforementioned studies illustrate that occupations are deeply affected by digital technologies as knowledge workers do or do not use these technologies to do their work, and demonstrate that workers may need to engage in identity work~\cite{Barrett1999-bv, Lamb2005-dh}. 

\subsection{GAI as a Digital Technology Augmenting Knowledge Work}
AI, and more recently, GAI specifically, challenge what digital technology means for occupations and their knowledge~\cite{brynjolfsson2023generative, DellAcqua2023-ia}. Due to its multi-modal, generative, and cross-domain applicability, GAI's economic potential has been explored in several knowledge-intensive domains such as consulting~\cite{DellAcqua2023-ia}, customer services~\cite{brynjolfsson2023generative}, and product development~\cite{Bouschery2023-gd}. Exploring knowledge workers' anticipations and expectations of GAI, \citet{woodruff2024knowledge} found that workers expected to outsource mundane tasks such as note-taking without forgoing any control and rejecting predictions of workforce automation through GAI. Similarly, workers in a large international technology company reported the use of GAI for supporting activities, such as creating work documents, generating new ideas, finding information, and improving their writing~\cite{brachman2024knowledge}. While the incorporation of AI might be initially aimed at supporting workers and enhancing productivity, a nascent research stream also points towards the unintended and unconsidered consequences of GAI. Power dynamics within organizations might shift as the introduction of AI in the workplace can result in heightened managerial oversight and a reduced valuation of the workers' practices~\cite{Monod2023-hf}. 

GAI, like previous digital technologies, can trigger transformations in not only what workers do but in how workers perceive their roles and their work. Opacity and complexity of new technology can deprive workers of the ability to understand and master the technology they are relying on as part of their work~\cite{Anthony2021-gr, Zednik2021-mh}. These trends are expected to be reinforced through GAI as it might feel like AI is executing work in their behalf. Literature illustrates the importance of studying workers' perspectives and the dynamics between workers within an organization to understand the transformations digital technologies can trigger, yet it is unclear how exactly GAI affects knowledge workers' occupational identity and which underlying factors can help explain workers' potentially varied reactions towards GAI. 

\section{RESEARCH CONTEXT AND SITE}
This study seeks to understand software engineers' sense-making of how GAI affects their occupational identity and their psychological needs. The case of software engineering is particularly compelling 
for two reasons. First, software engineering is a typical example of knowledge work where workers gain and apply analytical domain knowledge through formal training and extensive practice of, e.g., coding. Second, workers’ tacit and rich knowledge in the specific domain of software engineering enables them to potentially understand better the capabilities and limitations of GAI compared to knowledge workers from other, non-computational domains. 

To generate a grounded understanding of how GAI impacts knowledge work in the software engineering domain, 
we conducted an abductive study through survey data collection and semi-structured interviews with engineers at a software organization, which we refer to as SoftCloud~\cite{Schamber2000-mv,Hsieh2005-bh}. SoftCloud runs a monitoring platform for on-premises and cloud-based applications, thereby offering its clients services around forecasting, anomaly detection, and enterprise IT maintenance. 

Team members work together in preventing, identifying, and resolving issues clients have with their cloud IT systems and services. Junior engineers predominantly write scripts for debugging tickets or code user interface (UI) components for the monitoring platform. Senior engineers engage with the juniors by reviewing code and through team meetings, yet also engage in their own coding and research activities, e.g., by prototyping new features.  

The way in which coding tasks are performed in software engineering has evolved considerably through increasingly sophisticated technology and, more recently, through the commercial availability of GAI. GAI can potentially subsume many of the repetitive coding tasks historically performed by junior engineers. Further, while monitoring tasks still rely on the same vendors and their platforms, GAI, such as ChatGPT, can quickly retrieve and summarize such vendors’ documentation. 
Because software engineers require specific domain knowledge yet can potentially automate or outsource parts of their work with GAI, we propose that software engineering offers an ideal setting for theory building about how knowledge workers come to understand and rely on GAI, and with what underlying reasons and consequences.

\subsection{Data Collection}
This study was approved by an institutional review board at X university \footnote{Here and elsewhere in the manuscript, we remove any author-identifying information.}. Data collection was conducted remotely via online surveys and Zoom due to geographical dispersion and to ensure participants' privacy. For survey participation, participants were asked for written consent. For all interviews, the first author asked for verbal consent for participation, audio recording, and taking notes. The study phases and participants involved are summarized in Table~\ref{tab:data collection}.

\begin{table}[!ht]
\caption{Overview of data collection steps}
\begin{tabular}{{|p{4.9cm}|p{2.9cm}|}}
\hline
\textbf{Data Collection} & \textbf{Subjects} \\ \hline 
Phase 1: Informal Interviews & 2 executives, 1 junior
\\ \hline
Phase 2: Online Survey & 35 software engineers
\\ \hline
Phase 3: Semi-structured interviews & 11 software engineers
\\ \hline
\end{tabular}
 \label{tab:data collection}
\end{table}

\subsubsection{Understanding Knowledge Work: Informal Interviews}

First data collection for this study was informed by an unstructured review of online material and documentation available on social media platforms such as Reddit or Twitter and the promotional descriptions and data policies of commercial GAI vendors~\cite{Ess2004-jv}. Our initial work (December 2022 to March 2023) provided a foundation for understanding knowledge work augmentation through GAI and informed our case selection.

Subsequently, we conducted three informal interviews with knowledge workers from different domains: a junior employee working in consulting, a brand director working in marketing and creative services, and an engineering director working in software engineering.
As our goal was not to gain a comprehensive understanding of all types of knowledge work, but rather to learn whether discussions around GAI in news and media substantiated for actual knowledge workers in the field, these interviews provided valuable context. All interview partners shared the view that the use of GAI benefited their work in becoming more efficient or even automating certain sub-tasks. 

Interestingly, the engineering director mentioned that workers within the organization reacted differently to the new technology. Intrigued by the reported mixed reactions of workers within the same organization and the organization's interest in pursuing the implementation of GAI, our research team was given access to reach out to the engineers within the software organization. 
As we reached out to the engineers directly, management was not aware of who participated in our study. We communicated our role to all study participants, including the engineers and management, during the consent process, the distribution of the online survey, and the one-on-one interviews following approved institute review board guidelines.

\subsubsection{Understanding the Organization: Online Survey} 
To gather data in the context of one organization, we first solicited workers' work processes, key tasks, and views on GAI through an online survey~\cite{Miller2007-st}.
A recruitment email was sent to all employees in engineering-related positions, including management (May 2023 -- June 2023, 50\% response rate \footnote{The percentage of female workers is omitted for privacy reasons.}). 
The survey included both structured questions (i.e., 5-point Likert scale from `strongly disagree' to `strongly agree', e.g., `Using ChatGPT or another GAI tool improves my work.') and open-ended questions (e.g., `What 1-2 GAI tool(s) are you using most heavily or frequently? Please list the name of the tool(s) and a short description for which tasks you use them.'). Appendix~\ref{surveyresults} provides an overview of selected survey results.

\subsubsection{Understanding the Knowledge Worker: In-Depth Interviews} 
The survey results raised questions about the perceived impact of GAI on engineers’ work.
 We were particularly interested in the differences between GAI users and non-users, and engineers with varying levels of experience. 

We conducted eleven semi-structured interviews, lasting roughly 45 minutes (June 2023 -- August 2023). Interview questions tapped the participants’ current work practices and activities (e.g., ``How would you describe your role and tasks?''), general impressions and expectations around GAI (e.g., ``What do you find rewarding/concerning about GAI tools such as ChatGPT?''), and more concrete questions about participants' sense-making of GAI at work (e.g., ``Has ChatGPT replaced any tools you previously used or modified your work processes? If yes, how?''). Depending on participants' expressed use or non-use, we modified and extended questions to better understand the underlying reasons for their embrace or resistance to GAI (see Appendix~\ref{interviewguide} for our semi-structured interview guideline). Over time, we also reallocated the focus of our interview questions (i.e., we reduced the future-oriented questions on potential use cases of GAI within the organization towards more worker-focused questions of individuals' sense-making process when dealing with GAI for their work). 
Table~\ref{tab:participants} provides an overview of our participants who are pseudonymized for data documentation, analysis, and presentation. All interviews were conducted in English.
 Interview transcription was conducted by the first author, with the second and third authors reviewing transcripts for completeness and understandability. 

\begin{table}[!ht]
\caption{List of interview participants. S prefix in ID stands for ``senior''; J prefix stands for ``junior''. 
To not allow the identification of individual employees possible through a combination of multiple characteristics, we provide an aggregated overview of work focus. All employees work in engineering (i.e., software and/or user interface).
}
\begin{tabular}{|p{1cm}|p{4.3cm}|p{1.5cm}|}
\hline
\textbf{ID} & \textbf{Domain Expertise (in years)} & \textbf{GAI User}\\ \hline 
S1 & >12 & Yes
\\ \hline
S2 & >12 & No
\\ \hline
S3 & >5, <10 & Yes
\\ \hline
S4 & >5, <10 & Yes
\\ \hline
S5 & >5, <10 & Yes
\\ \hline
J1 & >3, <5 & Yes
\\ \hline
J2 & >3, <5 & Yes
\\ \hline
J3 & <3 & No
\\ \hline
J4 & <3 & Yes
\\ \hline
J5 & <3 & Yes
\\ \hline
J6 & <3 & No
\\ \hline
\end{tabular}
 \label{tab:participants}
\end{table}
\subsubsection{Data Analysis}
To analyze and evaluate the interview data, we employed a qualitative and thematic analysis~\cite{Braun2021-la, Rivard2012-vi, Hsieh2005-bh}. Our coding approach aimed to derive new insights and knowledge from a rigorous content analysis of our textual data. 
For the initial coding step, we followed an inductive content analysis~\cite{Saldana2014-gd, Schamber2000-mv}. The two coders developed codes independently and based on terms or phrases used by the knowledge workers in the transcripts. We did so to manage our preconceptions and to avoid being preliminary deceived by our kernel theory into overseeing subtle insights and interesting features early in the coding process. 
After discussing and comparing the codes, the two coders specified the rules of application on a common codebook and double-checked their coding against the common coding basis. Using our theoretical lens around occupational identity, we then reviewed codes and identified potential connections between concepts. Our theoretical lens was later extended to SDT. This structured analysis allowed us to generate clear definitions and names for our second-order concepts in line with our theoretical lens and to group our findings towards a more abstract, theoretical level. Both steps one and two were done by both coders in multiple iterations. In total, we conducted four rounds of coding. The second-order themes were discussed with the third author who provided feedback on the data analysis and connection to the theoretical lens of our study.

\section{RESULTS}
As part of the following, we explore our RQs centered around software engineers' sense-making of GAI for their occupational identity, and its impact on work-related psychological needs. 

\subsection{GAI's Influence on Software Engineers' Identity Work}
To better understand how software engineers made sense of the technological change induced by the introduction of GAI, we first briefly describe how software engineers framed their occupational identity. We then explore how engineers perceived GAI to impact their competence, autonomy, and relatedness. 

\subsubsection{Occupational Identity and GAI} \label{SEOccIdent}
When software engineers talked about their work and role in general, they often referred to the importance of domain experience and the distinctiveness of the engineering occupation. Workers relied on domain-specific vocabulary, such as abbreviations for programming languages. An illustration of the pride and distinction from other professions was given by J5:
\say{I think if you're going to get by it as an engineer, you can tell when someone's written the code and when someone has not written the code, and you can also tell when somebody's skills are progressing.}

Both junior and senior software engineers mentioned the necessity of domain expertise to execute their work successfully. 
The distinctiveness of software engineering from other occupations and the importance of domain expertise also became apparent when we asked about workers' general perception of GAI, as depicted by S2:
\say{It's really, really hard, because you need to have domain knowledge […] In our domain, sometimes you wouldn't even understand the answer [of GAI], just because you are not an expert.}

The software engineers also attributed their knowledge of GAI to their domain expertise and used their occupation as a signal of knowledge and rationale for making judgments about GAI. When being asked how they deal with oftentimes raised accuracy issues with GAI, S2 talked about how 
\say{as an engineer, I know this problem will get solved.}
A junior engineer, J4, delineated themselves from general, public opinions about GAI:
\say{I think that there are a lot of misconceptions about why [ChatGPT] is useful. But I definitely think it's exciting. [...] It's the first time that any new thing has come out since I've started my career as a software engineer.}

While we identified consistent themes around software engineers' occupational identity, we found that software engineers differed in their responses when being asked about how GAI affected their work. 
As summarized by one software engineer, S5, reactions to GAI seemed to be contingent on domain experience: 
\say{There's mixed reaction. Some of my teammates, especially the younger ones, are very pro ChatGPT, love it. Some of the older, [...], think it's just a parlour trick and it's all just trickery and it doesn't have much use in a way. 
So it's been a mixed reaction. And anecdotally, 
the more experienced part of the team is more cautious, shall we say.}
We were, thus, interested in better understanding how GAI impacted software engineers' understanding of their own role as well as the underlying reasons for potentially varied reactions.

\begin{table*}[t]
 \caption{Tensions of Competence and Related Patterns of Identity Protection}
\resizebox{\textwidth}{!}{
\begin{tabular}{p{4cm}||p{5.75cm}|p{5.75cm}}
 Domain Expertise & \textbf{Junior Software Engineers} & \textbf{Senior Software Engineers}\\
 \hline
 GAI's impact on engineers' \textbf{need for competence} & Efficiency gains in tension with potential lack of skill development & Limited productivity gains\\
 \hline
 Identity work to address threat to competence & - Recognizing the importance of developing own domain expertise &  -Recognizing concerns of uncontrollable data security issues\\
   & - Restraining use of GAI to mundane and limited tasks only & - Vieweing GAI as unsuitable to augment own work\\
   & - Treating GAI as a tool with implications similar to those of other digital technologies & - Recognizing the limitations of GAI\\
  \\
\end{tabular}
}
\label{tab:competence}
\end{table*}

\subsubsection{Need for Competence}  \label{Competence}
In many ways, software engineers saw GAI positively affecting their competence (an overview of GAI's impact on software engineers' need for competence and related identity work engineers engaged in can be found in Table~\ref{tab:competence}).
Software engineers highlighted the efficiency gains as a major benefit of using GAI, or using GAI when being stuck with a certain task, as illustrated by J5:
\say{It's just the speed. It's very fast. Let's pretend I'll send it a line of code that I don't understand [...] [GAI] saves you clicks, it saves you time typing which I think is every engineer's dream. It's just time that you're looking stuff up. I think just having that ability to reduce that time is very, very helpful.}
S3 further elaborated on the competence gains through GAI:
\say{It has overall increased my efficiency in the sense that when I am trying to work on something, previously something might have taken me a whole week to get up and running. Now with ChatGPT providing me with a shortcut to assimilating a lot of information and giving that to me---even considering the overhead of sifting through the bad or inaccurate responses---now I can, approximately, get that done in three, four days. It's not like I can do other work. It's more that I can do my regular work faster.}

However, the introduction of GAI also presented a major threat to juniors' need for competence. Junior engineers expressed their concerns about not developing their skills sufficiently. For example, J1 highlighted the importance of not always relying on GAI in order to develop their domain expertise:
\say{I try to only use [GAI] to speed things up. I think it could get dangerous if I don't take the time to think through something thoroughly, and see if I can figure it out on my own without that resource. If I just dive straight into that, I think it's taking away from expanding my brain in the way the way I want to learn things right. I want to know what my resources are, not just jump straight to ChatGPT [...] Even though it's a lot quicker to do that, I think it harms me in the long run.}
This quote shows tensions regarding the effect GAI has on software engineers' competence, especially for junior software engineers. The junior's hesitance can be viewed as a protection mechanism to develop skills independently and to maintain an (idealized) occupational identity.  

Interestingly, we did not find that senior engineers felt similarly threatened in their competence. Senior software engineers expressed few suitable scenarios in how GAI would augment their work, such as S2 responding to GAI by defending and strengthening their existing occupational identity:
\say{I have lots of feelings on it. I am now [xx] years old. I studied some AI back in college [...]. And everyone keeps saying `It's going to come'. [...] Especially with GAI, it can make new things. But when I talk about all the things that we look at our company [...] We don't need random things created.}
The quote depicts how the senior engineer stuck to their existing routines and viewed GAI as just another technology passing by but not directly impacting their work.

Next to recognizing the importance of establishing and developing one's own domain expertise, some junior software engineers also addressed the threat to competence by framing GAI as a tool, and by claiming the implications of GAI to be similar to those of other digital technologies. When being asked about their general thoughts about GAI, J4 replied:
\say{
It hasn't disrupted my daily flow or anything like that. It's a tool.} They continued: \say{I think there's nothing specific that I've been like `Wow! I can't believe it came up with that'. I'm more just like `Wow! It came up with what I would have gotten from Google but 20 minutes faster.' [...] It's just a tool.} 
J6 stressed the impact of GAI in a similar manner:
\say{I am hesitant about [GAI], I guess. I don't heavily lean one way or another. Obviously it's been in the news a lot, and a lot of people in the industry are talking about it and are excited about applicable uses. [...] I don't think it's as invaluable as some people make it out to be. At the end of the day, it's just another tool.}

An overview of GAI's impact on software engineers' need for competence and related identity work engineers engaged in can be found in Table~\ref{tab:competence}.


\begin{table*}[t]
 \caption{Tensions of Autonomy and Related Patterns of Identity Protection}
\resizebox{\textwidth}{!}{
\begin{tabular}{p{4cm}||p{5.75cm}|p{5.75cm}}
 Domain Expertise & \textbf{Junior Software Engineers} & \textbf{Senior Software Engineers}\\
 \hline
 GAI's impact on engineers' \textbf{need for autonomy} & Threat to ownership and agency & Threat to responsibility and control\\
 \hline
 Identity work to address threat to autonomy & - Stressing the importance of agency in relation to skill development &  -Stressing the importance of non-transferable accountability and monitoring of juniors\\
 & - Deflecting from oneself when thinking about the automation of junior positions & - Deflecting from oneself by identifying automation opportunities with service tasks and junior positions\\
\end{tabular}
}
\label{tab:autonomy}
\end{table*}

\subsubsection{Need for Autonomy} \label{Autonomy}
Software engineers expressed the importance of agency and independence of GAI as an augmenting technology. S4 said, \say{More or less, my attitude towards GAI is that I don't want to let the AI lead me around by the nose.}
The importance placed on notions of ownership and independence of GAI was particularly pronounced when talking to junior software engineers. J5 described the need for autonomy and how it was affected by GAI by describing their work: \say{It is what I do for 40 hours a week. You want to take some pride and have some ownership of it. And you don't want it to just all be spit out of the machine. [...] A lot of this job that I do is a lot of learning, and I try to learn something from it. So I feel like you're cutting yourself a little bit short when you're not doing these things.}

Senior engineers also felt their autonomy to be affected by GAI, yet for different reasons. Responsibility was a key aspect seniors referred to when discussing the impact of GAI. S2 expressed the importance of senior engineers' accountability and related threats of automation induced by GAI:
\say{If [GAI] gives the junior people bad advice, I still feel like senior people are gonna have to be watching the advice it gives and be like, ‘Oh, no, no, that's not the right advice.’}
They (S2) continued:
\say{[GAI] will boost the needle, but not in a way that it frees up a ton of time. Our senior people spend maybe 10 to 15\% of a year mentoring. And even if you drop that down to 5\%, you have to have 20 senior people to even reduce one [person].}
This quote is telling because it stresses the senior engineer's perspective on the remaining importance of managing junior workers and how their own work and role cannot be automated. At the same time, the quote points towards the idea that GAI potentially also influences workers' relatedness and how workers of different domain expertise are connected. Table~\ref{tab:autonomy} contrasts the different impact GAI had on juniors' and seniors' need for autonomy, respectively.

Software engineers also grappled with the impact of GAI on their autonomy by predominantly deflecting from themselves as a path to protecting their occupational identity. One junior engineer's (J6) quote illustrates this:
\say{I know some people ask ChatGPT to generate code or write scripts, which is, in fact, my entire job. But yeah, I guess a lot of my job is looking at pre-existing code or pre-existing libraries and then adding to it, so I can't really see a scenario where I would trust code generated by AI.}.
Another junior software engineer (J2) named entry-level engineers likely to be affected by GAI but did not consider their own job to be threatened:
\say{In terms of my own work, I think I'm concerned, not for myself, but for people who come after me, especially the younger entry-level people. Because from the little bits of code that I've had ChatGPT create personally, it does get about 95\% of the way. There are some basic mistakes that it makes but I think over time, it's pretty simple for it to solve it. [...] I think it's going to hurt a lot of the entry-level people when someone with more experience can instantly create the code that those people would have been creating previously.}
This quote ties back to the importance junior engineers placed on their skill development as being competent is directly tied to not becoming obsolete. 

Senior engineers also deflected from their own roles by identifying automation opportunities for customer support tasks and junior positions. S4 clearly distinguished themselves from junior engineers: \say{For what I'm doing. [...] I'm not having it actually write the functions that I'm doing. So for myself I'm not super worried about it.}
When asking about automation use case with GAI, engineers like S3 referred to service tasks: 
\say{For example, when we call customer support: [...] At first, you will have an AI chatbot that's trying to solve your problem. And if it doesn't, then you get to a human.}

\subsubsection{Need for Relatedness} \label{relatedness}
The rise of GAI appeared to create stark differences between junior and senior engineers at SoftCloud in how they saw their relatedness to colleagues and other engineers impacted by GAI. 
When being asked how colleagues and other engineers in the organization respond to GAI, S4 shared how they had to intervene juniors' use of GAI:
\say{
We had to snap down on people using Copilot pretty early on that we really can't have. Because we're writing proprietary code and so we can't have people shoving our proprietary code into a consumption engine so that everyone else can get to it.} 
Next to the issue of proprietary data, S4 added a second reason for their intervention: \say{We wanted to be sure that people weren't dumping a bunch of code that they didn't understand into the code base [...]. Ultimately, we want to make sure that people are writing their own code. Not because we think that the human code is necessarily better, but because if a human is writing it and understands it, they can go to take responsibility for it, and they can edit it when they need to.}
The senior engineer expressed their concerns about juniors' skill development. The reaction ties back to seniors' need for responsibility while also suggesting the necessity for junior engineers to take responsibility.

Senior engineers felt that they had to ensure that junior engineers were not using GAI in unintended ways that could potentially harm the organization as well as fear that the use of GAI can threaten junior engineers' skill development, as shared by S2: \say{The [junior] tech people have no idea what they're monitoring. They want to just tell their manager they're monitoring something, and if they're like, ‘Oh, this AI program told me I should monitor these things’, they're like, ‘That's fine’. But if it's not really what you should monitor, that's a problem.}

Junior engineers viewed themselves as more open and adaptive, especially as compared to more senior engineers who were viewed as more stubborn and fixed in their existing routines. Without being prompted, J1 differentiated between junior and senior engineers:
\say{I'm not stuck in my way, since I'm a junior, so I'm open to everything that's going to make my life quicker and me more efficient, and I know doing redundant things to me just takes away my time that I could be spending doing the things that ChatGPT can't do for us. [... ] I think there are [different opinions]. [...] I just think there's an old-school way of programming, and once you're kind of stuck in it, you're just stuck in it. It's just kind of tricky to convince people.}
When discussing the usefulness of GAI, J1 continued:
\say{It is helpful but I just think it can get dangerous in a sense of convenience, […] I just think it might be difficult to navigate in a team setting where one person is using ChatGPT and the other person's taking the old research method.} 
It becomes clear that the junior viewed this as a challenge to juniors and seniors working together. This common ground was perceived as particularly important and necessary for juniors' skill development. J3 described the importance of learning, and learning from fellow engineers, in this way: \say{There are benefits to having a human teacher. I think ChatGPT being primarily text-based... You don't get little emotions, you don't get little jokes here and there that create an experience [...] I think in long term learning, I think a human is still better because of those little things.}
Table~\ref{tab:related} provides an overview of GAI's impact on software engineers' relatedness and identity protection patterns particularly junior software engineers engaged in to preserve a common ground with the seniors.

\begin{table*}[t]
 \caption{Tensions of Relatedness and Related Patterns of Identity Protection}
\resizebox{\textwidth}{!}{
\begin{tabular}{p{4cm}||p{5.75cm}|p{5.75cm}}
 Domain Expertise & \textbf{Junior Software Engineers} & \textbf{Senior Software Engineers}\\
 \hline
 GAI's impact on engineers' \textbf{need for relatedness} & Loss of relatedness as seniors are viewed to stick to their existing work practices and routines & Loss of relatedness as juniors are viewed to use GAI in mindless and unintended ways\\
 \hline
 Identity work to address threat to relatedness & - Stressing the importance of learning from fellow engineers &  - Stressing the importance of monitoring and controlling junior work\\
  & - Focusing on developing skills and confirming with existing practices that enable a common ground with senior engineers &  \\
  \\
\end{tabular}
}
\label{tab:related}
\end{table*}

\subsection{Organizational Influences on Engineers' Identity Work}
\begin{figure*}[h]
\includegraphics[width=12cm]{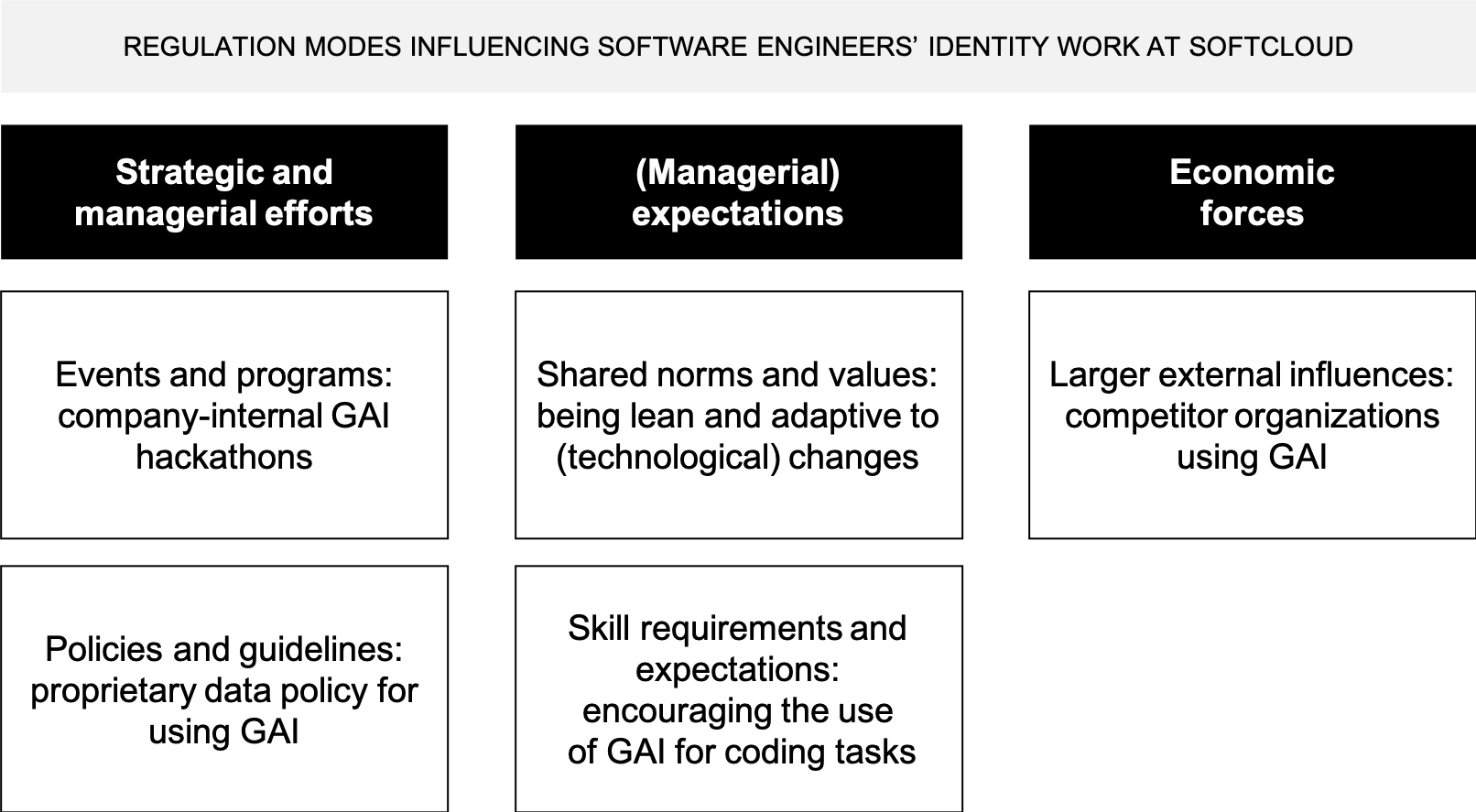}
\caption{\textsf{Overview of regulation modes of identity work at SoftCloud. Strategic and managerial efforts are organization-level initiatives that can indicate an organization's strategy. (Managerial) expectations position the organization in terms of its values and expectations toward its workers. Economic forces refer to larger organizational and economic forces that can transform 'the way of doing things' at an organization. This overview does not suggest to be en exhaustive presentation of possible regulation modes but rather an overview of the regulation modes that moderated the impact GAI had on software engineers' identity work.}}
    \label{tab:regmodes}
\end{figure*}

As mentioned earlier, the process of identity work can be conditioned by intra- and extra-organizational factors beyond individuals' sense-making, commonly referred to as regulation modes. In the context of our study, SoftCloud organized multiple company-internal hackathons on the theme of GAI. On the other hand, external influences on workers' sense-making can occur as a by-product of informal, unstandardized activities, e.g., due to larger social and organizational factors. Software engineers discussed GAI over coffee with their colleagues, for instance. In the following, we review regulation modes present at SoftCloud (for an overview, see Figure \ref{tab:regmodes}).

\subsubsection{Strategic and managerial efforts}
Strategic and managerial efforts are organization-level initiatives that can indicate an organization's strategy.

\paragraph{Company-internal GAI hackathons} 
Many software engineers talked about organization-wide hackathons on the theme of GAI, which they viewed as an encouragement of management to explore the use of GAI. As one junior engineer (J1) put it:
\say{It's still a meta opinion that our company is excited about AI. We had an internal hackathon where people tried to build something, using AI and that was cool. So I think the higher up of engineering is really intrigued with it, and ultimately they want to use it. There's no denying that it would make programmers faster. [...] They're gonna start using AI, and they're gonna start integrating it into whatever they're building.}
This framing is very contrasting to a junior engineer, J3, who did not use GAI:
\say{They put in a hackathon on the theme of GAI. So I think there is pressure from above to think of something new and cool that's gonna sell and sit apart from other tech companies. I think it's normal, for the hackathons but in the previous years it's been more like ``Just build something cool.'' But this year they wanted us to use GAI. That was the theme of both the hackathons I went to. So they definitely pressured us to use those tools.}

\paragraph{Proprietary data policy for using GAI} 
Engineers, e.g., S3, oftentimes referred to an internal policy that concerned the use of GAI for work:
\say{Our official policy is to not use ChatGPT for any official work. And I believe that's also driven particularly by privacy and security concerns. Typically, as far as I know, [name of organization] employees do not use ChatGPT for their actual day-to-day work.}
This policy was either used as a reason to not engage with GAI or to find a workaround to be still able to use GAI. A non-user of GAI, S2, made the first path quite clear:
\say{We had a statement from legal that said, `Do not put anything that's proprietary in ChatGPT.' Either customers' data, HR data, anything that's proprietary. If you can't say it to a human being outside of our company, you cannot say it to ChatGPT, which made it actually very clear. Because we go through extensive training on what you are and are not allowed to say to another human being outside of our company.}
On the other hand, some engineers stated that they did not share proprietary data with GAI yet found a workaround to still use ChatGPT. S4 described this approach, stating, \say{we had a policy come down from the top that said that we weren't allowed to just copy stuff out of ChatGPT and paste it into the code that we had to, you know, maturely modify it before we put it in} (S4). Other engineers mentioned that they used GAI for general, data-unrelated purposes or tasks outside of work at SoftCloud such as J1: \say{I actually use it a lot. I don't use it at work because we have regulations.}

\subsubsection{Managerial expectations}
Next to more explicit initiatives that expressed SoftCloud's strategic aim regarding the use of GAI, more implicit actions such as unspoken norms and codes of conduct equally influenced engineers' identity work. 

\paragraph{Shared norms and values} Software engineers at SoftCloud seemed to have a shared vocabulary, viewing the organization as agile and fast-moving, as described by S2: \say{A lot of the people on my team are new. So I have to coach them through learning the process, learning the tools, but then also how to do innovation. [...] We're a very lean organization. So you know how to do it efficiently.} Consequently, some engineers saw a natural, or even necessary, connection between the nature of the organization and the adoption of GAI, as the quote by J2 depicts: \say{I think it will definitely be a lot faster for us, we'll see the impact first with our company, because we're more dynamic and a younger company. So we can adapt a lot faster.} 

We observed that through latent messages from managers, an unspoken code of conduct was created around the exploration of GAI for work, as the following quote by S1 demonstrates:
\say{I think, companies that will resist the change, this adopting, they will have a lot of problems. [...] It's a cultural thing. So if you are culturing `I need to adapt, I need to change. My engineers can move and do something different.', it's awesome.}

Engineers were exposed to not only to different types of organizational influences but also conflicting and implicit messages from the top. The following quote by J4 summarizes the mixed signals that were sent through the policy and other efforts by the organization, as well as SoftCloud's attempt to actively influence engineers' use of GAI:
\say{I would say, one, my manager is very excited about all this. [...] He just has been encouraging the AI hackathons and he's always talking about ChatGPT. But organization-wide, we've just gotten a memo from legal that's like `Hey, be careful, you know. Don't copy and paste. Don't blame your eyes.' But there is unofficially, semi-officially, I guess, because it's coming from like upper levels of engineering, they're talking about it all the time, and definitely want us to use it. They want us to experience, or at least know a lot about it. I think you would seem `not in the know' if you didn't have opinions or thoughts, and how best to use it right now.} A senior engineer, S5, also mentioned that the use of GAI is steered by the top: \say{Well, it's coming right from the top, I think. It's not grassroot, I'm afraid. I think in the beginning it was. 
[...] And now it's coming from the top that we can start looking at it.}

\paragraph{Changing (engineering) skills} For both users and non-users of GAI, there was a strong shared understanding of the nature of their occupation. Their comments about this in relation to the organization-internal hackathon articulate the importance placed on domain expertise and engineering skills, as mirrored by user of GAI, J4: \say{We had a hackathon recently, and it was being led by this guy who's way smarter than me and had way more experience, and he made it in AWS and I'm trying to recreate it with the Azure components.}
J4 similarly expressed this appreciation of engineering skills in relation to the use of GAI, saying, \say{At the beginning, I found some really cool uses [of GAI]. [...] It was the week that ChatGPT dropped and I just made an account and used it. I was like `Okay, guys, you gotta check this out.' I had people come over and show them that. And that's probably happened two or three times since then.}
The strong valuation of engineering skills among the workers dominated workers' pride at SoftCloud. The use of GAI as a means to further express one's skills hereby acted as a mode of regulation as GAI was informally introduced into the organization. The expectation of using GAI was embraced by some of the engineers whereas others experienced the shift in expectations as pressure.

\subsubsection{Economic and competitive landscape}
Larger social, organizational, and economic forces provide a worker and an organization with particular conditions that can transform the ``way of doing things''~\cite{Alvesson2002-zr}. Such modes of regulations also triggered software engineers' identity work at SoftCloud.

A number of engineers, particularly users of GAI such as S3, shared this perspective on the need to evolve along large industry changes:
\say{I think in our industry as a whole, whenever something this transformative comes along, you have to evolve your own self alongside it, and so you have to adopt it into your regular workflow. [...] I feel like a lot of companies might feel pressured to start adopting GAI.} A junior engineer, J1, also mentioned the necessity to adopt GAI in order to not fall behind:
\say{So if we don't start considering that, we're behind, right? Because people already are doing it. [...] These are companies already incorporating it. And that's been going on for years. 
[...] 
}
This perspective seems to be also driven by engineers observing the market and competitors, as the following quote by S5 depicts: \say{There was a really good PR video by another company which works in the same sort of space [described how other organization is using GAI]. And that's how we would use it.}
J2 shared how management also recognized the competitive landscape when adopting GAI:
\say{I think the company will push entry-level people to explore using the AI more frequently, so it's sort of a hedging… just in case our competitors start using it and it is a huge competitive advantage}.
Non-users also shared this perspective yet distanced themselves from it, as the following quote by J3 depicts: \say{ I think the company wants to use [GAI] to stay relevant.[...] I mean, I'm not part of any management decisions or stuff. I'm just the developer, but it kind of feels that way.}
This quote illustrates how the engineer does not necessarily acknowledge the usefulness of GAI for the organization or themselves but rather describes it as a ``necessary evil''.


\section{DISCUSSION}
\begin{figure*}[h]
\includegraphics[width=12cm]{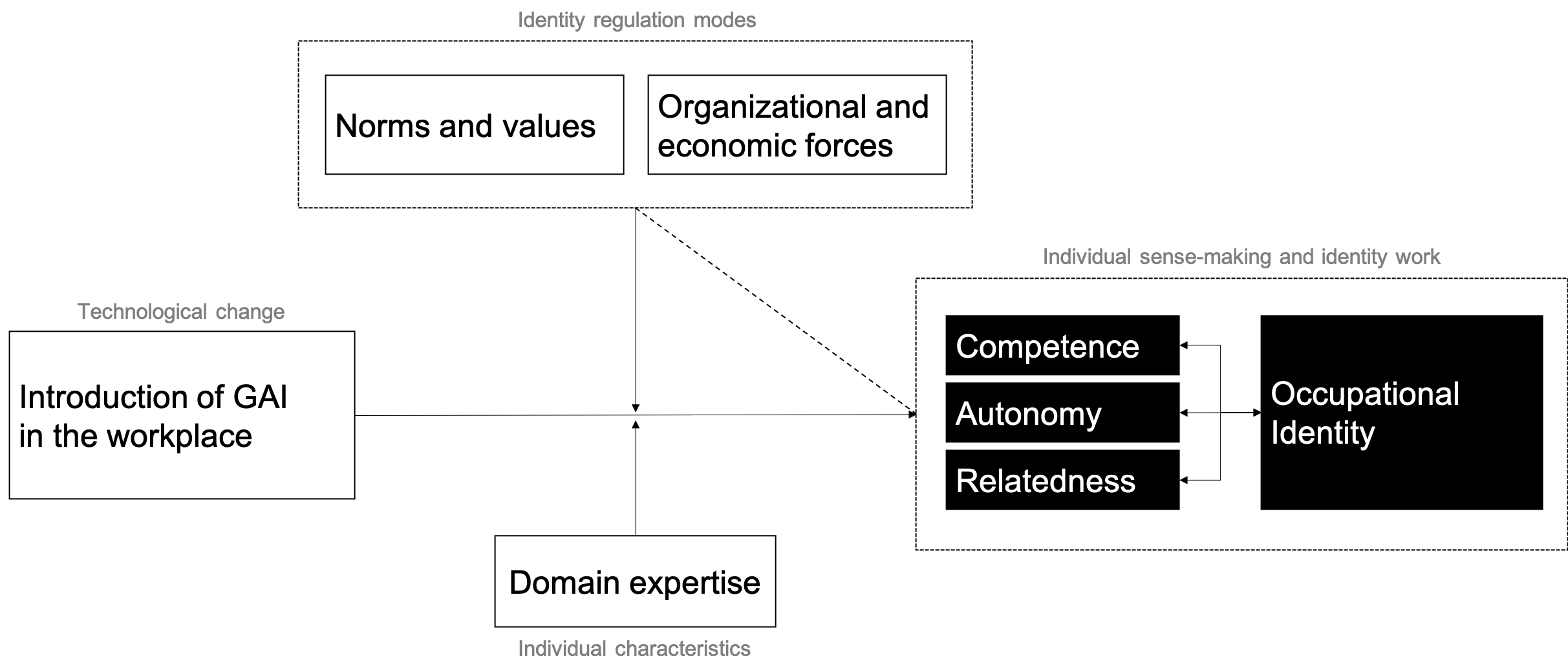}
\caption{\textsf{The impact of technological change on workers' occupational identity and key psychological needs. The conceptual model illustrates the effects of technological change on workers' sense-making of their occupational identity. Regulation modes and workers' domain expertise help explain these effects and can guide the impact of new technology. Highlighted in black is the individual sense-making of software engineers' identity, the main focus of our study's investigation.}}
    \label{fig:ds}
\end{figure*}

The adoption of GAI in organizations can be viewed as a crucial paradigm shift in the ever-evolving landscape of digital change in the workplace~\cite{burke2017organization}. 
The conceptual model in Figure~\ref{fig:ds} provides an overview of the key results of this study, i.e., software engineers' sense-making of their occupational identity in relation to the introduction of GAI in the workplace. Our conceptual model hereby builds on seminal notions of SDT and occupational identity~\cite{deci2012self,Gagne2022-lg, Ryan2000-vv} and illustrates our findings on i) how GAI impacted software engineers' need for autonomy, competence, and relatedness, and, in turn, ii) how software engineers engaged in identity work to protect and strengthen their occupational identity. We also identify how software engineers' identity work is further influenced by implicit organizational measures and external forces, commonly referred to as ``regulation modes''.  While we are exploring the impact of technological change in the workplace, we do so within the boundaries of the software engineering domain and at a specific point in time, i.e., during the rise and the adoption of the first commercially available GAI-based models. 

\subsection{Theoretical Implications}
Our findings highlight the relevance of occupational identity and human self-determination for understanding the impact of GAI for knowledge work in the specific context of software engineering. 

\subsubsection{Occupational identity and self-determination theory} 
\paragraph{Identity work} 
Our findings contribute to the body of literature exploring the (re)definition of occupational identity induced by the introduction of new technology~\cite{Beane2019-wi, Ibarra2010-xe, Nelson2014-kt, Vaast2021-ll}. Software engineers at SoftCloud varied in how they responded to the introduction of GAI. 
Software engineers had a strong occupational identity that is framed by their expertise and their distinction from other domains, which they also relied on to justify their assessment of GAI. Our study is congruent with earlier studies showing that highly skilled workers with established routines and identity may feel threatened due to the advent of a new technology~\cite{Vaast2021-ll} and might thereby, paradoxically, not fully leverage the potential benefits of these new technologies in order to preserve their existing occupational identity~\cite{Nelson2014-kt}. Literature has started to explore the underlying mechanisms of identity work, e.g., by better understanding the temporal dynamics of changing occupational identity~\cite{Vaast2021-ll}. As part of this study, we find that knowledge workers' level of domain expertise plays an important role in shaping identity work. Engineers' sense-making process served as a powerful mechanism to protect their occupational identity, enabling them to maintain their competence, autonomy, and relatedness in a changing professional landscape.  We hereby challenge prevalent assumptions of how GAI threatens learning and skill maintenance in knowledge organizations as we found that software engineers' patterns of identity protection enabled them to preserve and enhance domain-relevant skills and expertise. 

Illustrated along the three psychological needs of competence, autonomy, and relatedness, the findings of this study i) identify prevalent patterns of identity protection and ii) unpack the nuances of how GAI impacted junior and senior software engineers' sense-making of GAI for their work.

\paragraph{Competence} 
GAI's impact on software engineers' sense of competence created a tension between the appreciation of GAI as a useful tool to increase efficiency, and a worry about unsustainable over-reliance on GAI preventing the development of important domain skills. On the one hand, software engineers shared how using GAI helped them improve productivity and efficiency for selected tasks, such as replacing Google searches, thereby strengthening them in their perceived competence. At the same time, junior software engineers were concerned that an over-reliance on GAI could hinder necessary skill development. For senior engineers, implications for their competence were not as pronounced. Our findings illustrate how particularly junior engineers refrained from using GAI too much or for certain tasks in order to hone their domain-specific competencies and to maintain a common ground with their senior colleagues. While previous studies such as~\cite{Beane2019-wi, Anthony2021-gr} showed that the introduction of new technology (e.g., robots in medical surgery) can impair the skill development of more junior workers (e.g., surgeons), our findings call into question predominant assumptions of how the outsourcing of work to novel technology, i.e., GAI, might lead to a loss of knowledge and skills. To preserve their (idealized) occupational identity, junior software engineers' identity work served as a powerful mechanism to ensure the development of skills and preserve important tacit domain knowledge. 

\paragraph{Autonomy} Software engineers predominantly engaged in deflection mechanisms in response to GAI's impact on their autonomy. Junior engineers acknowledged automation gains, e.g., by outsourcing tedious and redundant tasks, yet deflected from their role when thinking about automation. Similarly, senior engineers acknowledged automation gains outside of their task area, e.g., by automating sales and service tasks and roles. Junior and senior engineers differed from one another in the coping mechanisms they engaged in to (re)claim their autonomy, which might be tied to their respective importance placed on notions of ownership and responsibility. While junior engineers expressed their desire for ownership and pride in their work, seniors' need for autonomy was expressed using responsibility and accountability. It was important for them to continue exercising their managing activities, including training the junior engineers of the organization. They expressed fears around junior workers training process, and their use of GAI in an unsupervised and unintended manner. 

\paragraph{Relatedness} 
Junior and senior engineers experienced and expected each other to behave and respond differently to GAI. This also led to discrepancies between juniors' and seniors' work practices, ultimately creating tensions of relatedness. Junior engineers experienced seniors to be more resistant to GAI and fixated on existing work practices and routines. These perceptions ultimately resulted in juniors' fear of losing common ground with senior engineers. Senior engineers experienced a loss of common ground with their more junior counterparts as they felt like junior engineers used GAI in a more reckless and unconsidered manner, and worried how such use would threaten juniors' skill development. Engineers desired to close the gap between juniors and seniors, i.e., expressed by juniors' importance placed on their skill development and preservation of domain expertise. At the same time, engineers feared widening the gap between juniors and seniors, i.e., expressed by seniors' (juniors') concern of junior (senior) engineers over-(under-)relying on GAI for their work.

\paragraph{Connecting occupational identity and SDT} 
In addition, our findings illustrate how the explicit consideration of domain expertise offers a fresh yet important perspective on extending our understanding of occupational identity and identity work. This perspective is enabled by integrating ideas of occupational identity with alternative theoretical frames, namely SDT. While work on SDT predominantly focuses on the role of self-determination for motivation, our work contributes to a richer understanding of the relevance of self-determination and its associated psychological needs of competence, autonomy, and relatedness for identity in organizational settings. Our study illustrates how threats regarding workers' psychological needs for self-determination can trigger important identity work. 

\subsubsection{Modes of regulation} 
Lastly, we examine the effect of conflicting organizational measures on engineers' identity work. Our interviews demonstrate how conflicted some of the engineers' opinions on GAI were. This cautious appreciation of GAI and mixed feelings might be driven by a general ambiguity and constant developments accompanying GAI, but also by the tension created by internal policies and conflicting information and measures driving and hindering the use of GAI. While not giving software engineers explicit orders or directions, changing values and new managerial expectations regarding GAI were consciously enacted as a form of organizational control. This became noticeable through the internal hackathons that were formally organized by SoftCloud, as well as expectations and latent messages shared by top management. Our study thereby also contributes to the body of literature on identity regulation. 

SoftCloud's norms and values acted as modes of regulation that triggered identity work and hence further impacted workers' occupational identity.  This is consistent with findings in the literature that indicate that organizational interventions and other external influences can impact workers' occupational identity and identity work~\cite{Alvesson2002-zr}.
While we previously saw that workers informally introduced GAI into the workplace, management attempted to further influence and increasingly steer workers' use of GAI. We found that the shared organizational values of SoftCloud as a competitive, lean company, as well as expectations regarding workers' capabilities, induced the engineers, particularly juniors, to test and experiment with GAI. As such, the rise of GAI created new or modified expectations about what software engineers are capable of doing. However, by some engineers, especially the non-users of GAI, the enforced expectations around GAI and activities like the hackathon were experienced as enforced, top-down pressure.

Larger organizational and economic forces seemed to influence workers' sense-making of GAI, such as competitors already deploying GAI, and thereby incite pressure on engineers to ``not fall behind''. This is in line with the notion of ``spillovers from technological change'', suggesting that changes in the larger, institutional environment (e.g., related occupations in the same domain) can enforce a need to adapt with individual workers~\cite[p. 609]{Bechky2020-nd}. These observation strengthens the notion of occupational identity being in constant flux~\cite{Vaast2021-ll}. 

Congruent with identity literature, our results show that these external forces and organizational measures exist in tension with one another, ultimately enabling a multiplicity of identity work practices. In other words, engineers regulated the introduction of GAI differently, e.g., by actively embracing or passively accepting GAI, despite being exposed to the same or similar organizational interventions. This finding induces interesting questions about what extent of occupational identity and identity work i) are formed by personal, individual practices versus organizational measures, and ii) can be deliberately shaped versus implicitly influenced.

\subsection{Design Guidelines for Identity Work}
In our study, workers informally introduced and used GAI for their work, as well as refrained from using GAI despite the encouragement of management to use it. We saw that it was important for engineers to preserve their occupational identity so that their needs for competence, autonomy, and relatedness were fulfilled (and not threatened). Research exploring the impact of technological innovations on the nature of work has raised the question of ``whether and how technology designed for other purposes [...] can be deliberately designed to met these core human needs [of SDT and] what can be done to influence the process to create more human-centered designs''~\cite[p. 388]{Gagne2022-lg}. Organizations, management, as well as system designers should be aware of these identity-protecting sense-making mechanisms as deployed technology cannot be expected to be simply adopted. More so, these stakeholders can take an active role in maintaining tacit knowledge within an organization and improving the skills and domain expertise of their workers by nurturing workers' identity work.

\subsubsection{Moderating technological changes through strategic measures}
As organizations are invested in technological changes having a positive impact on workers' productivity yet also self-fulfilment, it is crucial for organizations to proactively co-design the adoption and the use of the new technology. This becomes particularly important given the ambiguity and uncertainty accompanying increasing ``General Purpose Technologies'' such as GAI~\cite{eloundou2023gpts, Goldfarb2023ML}. The rise of GAI has shown that novel technology no longer is expected to be formally introduced and controlled by the organization yet can be informally introduced by workers themselves. It therefore becomes even more important for organizational leaders to design the adoption of technology in a proactive and transparent manner. Organizational measures hereby become relevant to steer and facilitate the use of novel technology to offer workers a common ground and to reduce any ambiguity that accompanies the use of the technology within the organization. The adoption of a new technology, its processes and tasks, can be facilitated in a variety of ways. The strategic and managerial level of an organization is hereby key in order to define a strategy regarding, e.g., the adoption of GAI, to specify the strategic objectives surrounding this adoption~\cite{Barricketal2015}. This can be done by explicitly delineating the goals but also the managerial expectations of workers using the new technology. While not all workers feel the need to adopt a disruptive technology, coherent policies and strategic measures can encourage the adoption. The implementation of new technology in work processes can also be designed in a participatory way so that the adoption of a technology is not experineced as top-down only. As an example from SoftCloud, workers realized that management was strategically interested in pursuing the use of GAI, e.g., by having their hackathons focus on the topic of GAI and thereby also offering engineers to take part in the conversation of how and for which tasks GAI should be used at SoftCloud. However, the direction and the explicit expectations of management towards the workers was not very clear as engineers also voiced ambivalence regarding the use of GAI and pressure from top management. 

\subsubsection{Designing for meaningful work}
Beyond an organization's strategic direction of technology adoption, crafting meaningful work, i.e., in terms of experiences, interactions, and tasks for an individual worker, can be initiated by top management~\cite{humprheyetal2007}. Different from strategic measures, work design can but does not have to directly refer to a novel technology. Our interviews suggested that while some engineers leveraged their existing skills and developed new skills using GAI, others explicitly refrained from using GAI---either to preserve their existing work practices or as a means to ensure the development of specific skills. Organizations are also required to be sensitive to workers' fears and judgement. Some of the engineers at SoftCloud mentioned having little to no exchange with their colleagues regarding the use of GAI, and others even expressed concerns regarding the quality and scope of interaction with and feedback from others. Forms of regulation can be used to empower workers, e.g., by providing additional opportunities to interact with others, or additional work programs concerned with the re- and upskilling of workers. These programs and trainings can but do not have to pertain to GAI-specific skills. But work design could also be co-enacted and changed by workers themselves. Research has found that proactively crafting their work, e.g., in terms of skills used and learned, as well as leveraging disruptive events induced by novel technology, can have positive implications on workers' sense-making and work environment~\cite{hall2002careers}. Software engineers mentioned that the rise of GAI required them to flexibly adapt to the changes this technology would induce for their work. As workers felt threatened in their occupational identity in response to the rise of GAI, work design and occupational identity are expected to be mutually influencing each other.

\subsubsection{Considering workers' domain expertise}
Feedback and quality interactions across rank levels and within teams are expected to be more important than ever. Our interviews illustrated how domain expertise moderated engineers' sense-making process and engineers' perceptions of being able to hone their skills and the outcomes of their work. Considering junior engineers' fears of not fully developing their skills in early phases of their job and thereby losing touch with more senior engineers, organizational efforts should pay particular attention to workers with less domain expertise. These efforts can entail the provision of sufficient development opportunities yet also vast interaction opportunities with senior workers within and outside of the team, similar to the hackathons SoftCloud organized. Junior engineers reflected on the fact that writing code on their own was necessary for them to gain the needed experience, whereas senior engineers voiced concerns about juniors using GAI in unforeseen or unintended ways, i.e., by learning from nonsensical information or from unreliable sources. Dedicated interaction opportunities would not only enable juniors to obtain the necessary skills and experience, yet could reduce bias seniors have about juniors' technology use (and vice versa).

Taking our proposed design guidelines and existing frameworks on work design~\cite{Gagne2022-lg} as a point of departure, we urge future research to explore how modes of regulation could be deployed to regulate workers' identity work.

\subsection{Limitations and Future Research}
Our findings should be interpreted in the light of certain boundary conditions, pointing towards important avenues for future research. It is important to note that our study captures a specific moment in time, i.e., the early adoption of GAI, as part of which we explored the dynamics of how people deal with change induced by novel technology. Due to the ever-changing and transient nature of GAI, our study by no means offers a holistic understanding of the impact of GAI but rather provides a specific snapshot in time. Due to the nature of our single case study, we focused on the domain of software engineering, a domain that is likely to be strongly affected by GAI due to its proximity to and reliance on digital technologies. While previous work studying occupational identity affected by new digital technologies considered comparable domains~\cite{Anthony2021-gr, Vaast2021-ll}, the question arises of how our findings can be generalized to other domains of knowledge work, as well as other types of work. 

Taking our informal conversation with an executive in the creative industry as a point of departure, it would be interesting to better understand how GAI affects knowledge work that hinges on creativity, novel ideas, and generativity. While GAI has been claimed to offer emerging capabilities that appear after the initial design, more recent studies have pointed towards the idea that these emerging capabilities are limited or even non-existent~\cite{Schaeffer2023-fa}. In a different vein, work on HCI has considered how blue-collar work is affected by novel digital technologies~\cite{Sheehan2023-ba}. As values and norms of factory workers are shaped by the introduction of computing services, future research would be useful to better understand the impact of GAI on workers' occupational identity beyond that of knowledge work.

Extant work exploring the impact of new digital technologies in the workplace have also shed light on distinct cultures and regions beyond the dominant understanding of how Western countries deal with technology~\cite{karusala2023speculating}. First studies have pointed towards harmful bias and undesirable stereotypes in the output of GAI~\cite{Abid2021-iw, Ganguli2022-oy, Nadeem2020-lv}. Another alternative research design might compare knowledge workers with similar occupations across cultures and regions, offering a previously neglected yet important perspective into how workers deal with stereotypes, bias, and discrimination in the context of occupational identity.

\section{CONCLUSION}
Our qualitative interview study with junior and senior software engineers in a medium-sized software company enables us to observe engineers' sense-making during the early days of the introduction on GAI, and to shed light on the  conflicting forces and psychological needs of engineers within the organization. We report how engineers' occupational identity is threatened differently by GAI, and how the engineers deal with these threats depending on their domain expertise. We discuss how the introduction of GAI is moderated by formal and informal forces in the workplace. We outline how organizations might regulate the introduction of GAI more consciously to enable a desirable and sustainable augmentation of knowledge work through GAI as part of which workers feel strengthened in their occupational identity.


\bibliographystyle{ACM-Reference-Format}


\appendix
\section{Survey Results} \label{surveyresults}

\begin{table}[!ht]
\caption{Exemplary Job Titles of Junior and Senior Engineers}
\begin{tabular}{{|p{1.5cm}|p{0.5cm}|p{5.4cm}|}}
 \hline
 Domain Expertise& N&Exemplary Job Title\\ [0.5ex] 
 \hline\hline
 \hline
 Junior& 11&Software Engineer, Software Developer, Software Architect\\
 Senior& 17& Director of Engineering, Senior Software Engineer, Manager\\
 \hline
 \end{tabular}
 \label{tab:junsen}
\end{table}

\begin{table}[!ht]
 \caption{Frequency of GAI Use of Junior and Senior Engineers}
\begin{tabular}{{|p{1.5cm}|p{2.5cm}|p{1.4cm}|p{1.4cm}|}}
 \hline
 GAI Use& Frequency (\%)&Junior&Senior\\ [0.5ex] 
 \hline\hline
 \hline
 Never& 13 (46.4\%)&6&7\\
 Rarely& 7 (25.0\%)&3&4\\
 Sometimes& 6 (21.4\%)&2&4\\
 Frequently& 2 (7.1\%)&0&2\\
 \hline
 \end{tabular}
\label{tab:freqGAI}
\end{table}

\begin{table*}[h!]
 \caption{Types of GAI Tools Used by Junior and Senior Engineers}
\begin{tabular}{ |p{3cm}||p{1.5cm}|p{1.5cm}|p{1.5cm}|p{1.5cm}|p{1.5cm}|p{1.5cm}|  }
 \hline
 \multicolumn{7}{|c|}{To what extent are you using (or have you used) the following GAI tools before?} \\
 \hline
 Type of GAI Tool& Never used before&Tried once or twice& < Once a week & 1-3 times a week & Once a day & > Once a day\\
 \hline
 ChatGPT& - & 5 & 3 & 5 & 1 & 1\\
 BART& - & 15 & - & - & - & -\\
 Midjourney& 12 & 3 & - & - & - & -\\
 Bard& 12 & 1 & - & 1 & 1 & -\\
 Dall-E& 12 & 1 & 1 & 1 & - & -\\
 Stable Diffusion& 13 & 1 & - & - & - & 1\\
 \hline
\end{tabular}
\label{tab:typeGAI}
\end{table*}

\begin{table*}[h!]
\caption{Perceived Risks and Concerns of Using GAI of Junior and Senior Engineers}
\begin{tabular}{ |p{5.3cm}||p{1.4cm}|p{1.4cm}|p{1.4cm}|p{1.4cm}|p{1.4cm}|  }
 \hline
 \multicolumn{6}{|c|}{How true are the following concerns for you when thinking } \\
 \multicolumn{6}{|c|}{of the use of ChatGPT or a similar tool at work?} \\
 \hline
 I worry that by using ChatGPT or another GAI tool...& Strongly disagree&Somewhat disagree& Neither agree nor disagree& Somewhat agree & Strongly agree\\
 \hline
 ... my work becomes less authentic.& 1 (4) & 1 (6) & 2 (2) & 6 (1) & 1 (2)\\
 ... no one can be held accountable if the information provided by the tool is wrong.& 3 (2) & 3 (5) & 0 (1) & 0 (3) & 5 (4)\\
 ... my colleagues view me as less skilled or proficient if they knew I used such a tool.& 2 (3) & 5 (5) & 3 (2) & 1 (5) & -)\\
 ... my superiors view me as less skilled or proficient if they knew I used such a tool.& 2 (3) & 4 (5) & 4 (1) & 1 (6) & -)\\
 ... I view myself as less skilled or proficient.& 1 (5) & 5 (6) & 3 (0) & 1 (1) & 1 (3))\\
 ... the information or data I submit to such tool is used in a way I did not foresee.& 0 (2) & 0 (1) & 0 (2) & 3 (4) & 8 (6))\\
 ... the information or data I submit to such tool is shared and used by some other party.& - & 0 (1) & 0 (1) & 3 (2) & 8 (11))\\
 \hline
 \multicolumn{6}{|c|}{Note: Numbers indicated in junior (senior) respectively.} \\
 \multicolumn{6}{|c|}{Survey participants were not required to respond to all questions.} \\
 \hline
\end{tabular}
 \label{tab:concerns}
\end{table*}

\begin{table*}[h!]
\caption{Perceived Benefits of Using GAI of Junior and Senior Engineers}
\begin{tabular}{ |p{4.5cm}||p{1.4cm}|p{1.4cm}|p{1.4cm}|p{1.4cm}|p{1.4cm}|  }
 \hline
 \multicolumn{6}{|c|}{How true are the following concerns for you when thinking } \\
 \multicolumn{6}{|c|}{of the use of ChatGPT or a similar tool at work?} \\
 \hline
 Using ChatGPT or another GAI tool...& Strongly disagree&Somewhat disagree& Neither agree nor disagree& Somewhat agree & Strongly agree\\
 \hline
 ... improves my work.& 1 (1) & 1 (1) & 1 (1) & 2 (3) & 0 (4)\\
 ... gives me confidence in my work.& 2 (1) & 0 (3) & 1 (2) & 2 (2) & 0 (2)\\
 ... does not really help me with improving the quality of my work.& 0 (1) & 2 (3) & 0 (2) & 2 (1) & 1 (3)\\
 ... does not really help me with improving the creativity of my work.& 1 (1) & 1 (2) & 0 (2) & 0 (2) & 2 (2)\\
 ... does not really help me with improving the creativity of my work.& 0 (2) & 1 (5) & - & 3 (0) & 1 (3))\\
 \hline
 \multicolumn{6}{|c|}{Note: Questions only posed to users (N = 15). Numbers indicated in junior} \\
 \multicolumn{6}{|c|}{(senior) respectively. Survey participants were not required to respond to all questions.} \\
 \hline
\end{tabular}
 \label{tab:experiences}
\end{table*}

\section{Interview Guideline} \label{interviewguide}
Current Work Practices
\begin{itemize}
\item To get a better idea of your role and your work: How would you describe your role and related tasks?
\item Think about yesterday: Can you walk me through your key work activities; from entering the office / opening your laptop to leaving the office / closing your laptop?
\end{itemize}
General Impressions and Expectations of GAI
\begin{itemize}
\item What do you know about ChatGPT? Are there aspects about ChatGPT you are unsure about / you would like to know more about?
\item What do you find rewarding about GAI tools such as ChatGPT in today's day and age?
\item What do you find concerning about GAI tools such as ChatGPT in today's day and age? [e.g., why would you not use it for your work] [Followup: Tell me why this is a concern] [Followup: Are there any privacy-related concerns?]
\end{itemize}
GAI Use
\begin{itemize}
\item Do you use GAI-based support tools such as ChatGPT for your work? If yes, what kind of tools do you use?
\item Do you have any expectations around the use of ChatGPT within your organization / within your work practices? If yes, what are some of these expectations?
\item Does your organization do anything to enforce these expectations?
\item Do you experience any ChatGPT-related discussions within your organization? If yes, can you give an example of a discussion you faced?
\item Can you think of a recent situation where you used ChatGPT for work and walk me through it?
\item Once ChatGPT generates an output to your request, how do you integrate it into your existing task / work flow? 
\item On what occasions do you find the AI to be useful? Why? (What's working for you in regards to ChatGPT?) [if follow up: What's the best piece of work you've gotten from ChatGPT?]
\item On what occasions do you find the AI to be not useful? Why? (What's not working?)
\item Have you changed the way you perform tasks? Which ones and how?
item Would you carry out the same tasks without the use of ChatGPT again? Why / why not?
\end{itemize}
Envisioning Future Use of GAI
\begin{itemize}
\item Has the use of ChatGPT created additional tasks or work for you? Do you think it will create additional or new tasks or work for you in the future?
\item Has the use of ChatGPT enabled you to take on other work-related acitivities (that you were previously not been able to do)? Do you think it will enable you to do so in the future? If yes, what kind of activities?
\item  How do you think would GAI potentially influence the structure and roles (within the organization)?
\end{itemize}
Closing Remarks
\begin{itemize}
\item Do you want to share any other thoughts or comments you have regarding ChatGPT? Did we forget an important aspect of your work?
\item Do you have any questions regarding this research project or interview?
\end{itemize}

\end{document}